%% file: sample-manuscript.tex
\documentclass[manuscript,screen]{acmart}

\AtBeginDocument{%
  }

\setcopyright{acmlicensed}
\copyrightyear{2025}
\acmYear{2025}
\acmDOI{XXXXXXX.XXXXXXX}
\acmConference[Conference acronym 'XX]{Proc. ACM Interact. Mob. Wearable Ubiquitous Technol.}{June 03--05,
  2025}{Woodstock, NY}
\acmJournal{IMWUT} 
\acmISBN{978-1-4503-XXXX-X/2018/06}

\begin{document}

\title{DuoZone: A User-Centric, LLM-Guided Mixed-Initiative XR Window Management System}

\author{Jing Qian}
\authornote{Both authors contributed equally to this research.}
\email{jq2267@nyu.edu}
\affiliation{%
  \institution{VIDA, New York University}
  \city{Brooklyn}
  \state{New York}
  \country{United States}
}

\author{George X. Wang}
\authornotemark[1]
\email{xw3617@nyu.edu}
\affiliation{%
  \institution{New York University}
  \city{Brooklyn}
  \state{New York}
  \country{United States}
}

\author{Xiangyu Li}
\affiliation{%
  \institution{Department of Computer Science, Brown University}
  \city{Providence}
  \state{Rhode Island}
  \country{United States}
}

\author{Yunge Wen}
\affiliation{%
  \institution{New York University}
  \city{New York}
  \state{New York}
  \country{United States}
}

\author{Guande Wu}
\email{guandewu@nyu.edu}
\affiliation{%
  \institution{Tandon School of Engineering, New York University}
  \city{New York City}
  \state{New York}
  \country{United States}
}

\author{Sonia Castelo Quispe}
\affiliation{%
  \institution{Visualization and Data Analytics Lab, New York University}
  \city{New York}
  \state{New York}
  \country{United States}
}

\author{Fumeng Yang}
\email{fy@umd.edu}
\affiliation{%
  \institution{Department of Computer Science, University of Maryland College Park}
  \city{College Park}
  \state{Maryland}
  \country{United States}
}

\author{Claudio Silva}
\affiliation{%
  \institution{New York University}
  \city{New York City}
  \state{New York}
  \country{United States}
}

\renewcommand{\shortauthors}{Trovato et al.}

\begin{abstract}
  \input{sections/abstract.tex}
\end{abstract}



\received{20 February 2007}
\received[revised]{12 March 2009}
\received[accepted]{5 June 2009}

\definecolor{fycolor}{HTML}{7765db}
\newif\ifnotes
\notesfalse 
\newcommand{\fy}[1]{\ifnotes{\leavevmode\color{fycolor}{(Fumeng: #1)}}\fi}

\maketitle

\input{sections/01_introduction.tex}

\input{sections/02_related.tex}

\input{sections/03_system}

\input{sections/04_user_study}

\input{sections/05_discussion}

\input{sections/06_limitation_n_future}
\input{sections/07_conclusion}
\input{sections/08_acknowledgement}
\bibliographystyle{acm}
\bibliography{references, sample-base}

\end{document}

%% file: sections/abstract.tex
Extended Reality (XR) offers portable, private, and extensive workspaces by extending virtual displays beyond desktops. Current XR systems provide users with limited functionality to manually place, arrange, and resize virtual windows for productivity work, leading to increased interaction costs and fatigue over time. We propose a mixed-initiative, human–AI window management system that preserves users' control and agency while automating support for efficient window management. Through user-created spatial zones, users can swiftly perform virtual window management while a large language model (LLM) provides further hints for multi-scenario efficiency (e.g., work, study, entertainment) and suggestions for low interaction costs. A user study (N=16) with Apple's Vision Pro found that our proposed spatial zones lead to significantly faster adjustments and lowered effort and cognitive load compared to the existing baseline, while the LLM's feedback supports better scalability and fulfills the potential for future automatic virtual window management.

%% file: sections/01_introduction.tex
\section{Introduction}
\begin{figure}
\centering
    \includegraphics[width=0.85\textwidth]{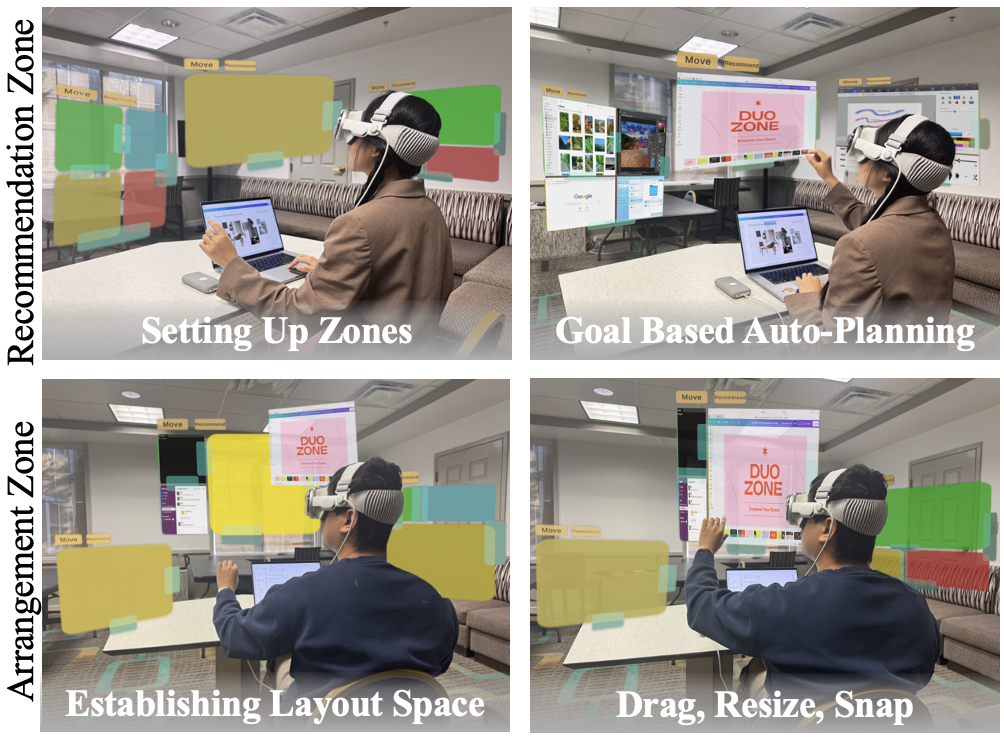}
    \caption{DuoZone uses two spatial configurations (or zones) to achieve efficient and low cognitive load XR window management via human-AI collaboration. A Recommendation Zone (top row) offers users spatial layouts to establish areas of interest for window management swiftly. Based on established layouts, the system automatically recommends relevant applications and adjusts layouts based on users' high-level goals conveyed through voice interaction or text input. In the Arrangement Zone (bottom row), users refine the spatial layout by establishing layout space and then arranging applications using dragging, resizing, and snapping.  }
    \Description{Teaser}
    \label{fig:teaser} 
\end{figure}
The emergence of high-compute power, light-weight, and high-quality extended reality (XR) systems has opened a new paradigm of portable work to free users from physical screens' constraints~\cite{pavanatto2021we}. These XR devices offer a large interaction canvas with privacy, allowing the ultimate portability to have an accustomed workspace adapt to mobile working~\cite{schmalstieg2000bridging,pavanatto2024multiple} and letting users handle comprehensive and complex work outside the conventional office~\cite{10.1145/3586183.3606717, pavanatto2025spatial}. One crucial benefit of XR working is that it provides an ad-hoc multi-window experience,  allowing users to multitask and work on a desktop without carrying multiple physical screens. 

However, existing research found that three main interaction challenges remain for XR working. First, even simple actions such as moving or resizing a virtual window can be challenging in XR~\cite{wang2025handows}, causing more effort and physical load on users over time. This becomes more challenging when users attempt to set up their workspace or create a layout in XR. Second, there is a lack of an automated way for XR window management. The current way to set up a workspace or manipulate virtual windows still relies on fully manual adjustment in the state-of-the-art XR devices. 
Third, view occlusion caused by a virtual window can undermine the effectiveness of XR applications~\cite{10.1145/3654777.3676470}, not only raising safety concerns but also reducing users’ ability to perceive and process both digital and physical information.

As a result, despite the known benefits, current XR systems for productivity and work suffer from interaction inefficiencies that increase interaction time, effort, and cognitive load. Optimizing virtual window interaction is therefore pivotal for enabling users to focus on their tasks and enhancing overall XR work productivity. This raises an important question: How can we design and implement an interaction system that improves user performance while reducing effort in virtual window management?


We propose \textit{DuoZone}, a mixed-initiative, human-AI collaborative virtual window management system aiming to increase interaction efficiency, reduce time and cost in adjusting layouts, and explore automatic window management (Figure~\ref{fig:teaser}). The virtual windows described in this paper are not AR labels on which most of the existing work focuses~\cite{bell2001view,grasset2012image}; rather, they are interfaces such as a browser, IDE, or terminal that users use for work and productivity. The key innovation is to allow users to create two types of interactive \textit{zones} in XR, which can be used by both an LLM and the user to adjust their position, scale, layout, and ordering. While an \textit{arrangement zone (first type)} offers functions such as snapping and template-based resizing for users to swiftly adjust the virtual windows, the \textit{recommendation zone (second type)} serves as channels for AI to fill in the zone with task-relevant virtual window applications and adjust the layout for reduced effort and interaction speed. 

To understand and explore the advantages of automated window management and how it can enhance user experience, we conducted a user study (N=16) with design and engineering experts who have at least 3 years of professional experience. We found that arrangement zones significantly improve speed, reduce interaction time, and cognitive load compared to the existing manual window arrangement. Recommendation zones further significantly improved the speed of creating XR workspaces and lowered the cognitive load and effort for two different scenarios (design and programming), with users accepting more than $90\%$ of AI's app recommendations, $76.5\%$ of AI app ordering, and $82.8\%$ of layout suggestions. Meanwhile, participants demonstrated an acceptable workflow where user-created zones filled by AI recommendations, followed by fine-tuning in size and application type. Overall, participants using the DuoZone system demonstrated higher performance and lower effort, and positive on the collaborative approach to retain agency of their window management. 

Our contribution is two-fold, and the system is open-sourced on [link to provide upon acceptance]:

\begin{enumerate}
    \item The DuoZone system that uses two types of configurations to support swift window layout management and human-AI collaborative layout management.
     \item An empirical study revealed that 1) the arrangement zone improves window adjustment, sizing, and layout efficiency while lowering the operation effort and cognitive load; and 2) the recommendation zone provides a highly accepted human-AI collaborative workflow that shortens the setup time and lowers effort in XR workspace setup.
\end{enumerate}

%% file: sections/02_related.tex

\section{Related Work}
\subsection{Multi-window management in AR/VR}
Window management is a system interface that controls how users view, organize, and interact with on-screen applications by managing display resources and directing input to the right program~\cite{robertson2000task, beaudouin2001novel, lee2017projective}. 
Effective window management supports task management by enabling users to organize, resize, and switch between multiple windows efficiently~\cite{wang2025handows,robertson2000task}. In AR/VR, virtual window management is more complex than traditional 2D counterpart because these windows can float freely within immersive environments and attach to various spatial reference frames such as the world, body, or user’s view~\cite{wang2025handows, ens2014personal, pei2024ui}. While this extra freedom enables expanded workspace, contextual multitasking, and more natural spatial cognition~\cite{pavanatto2024multiple, pavanatto2021we, pavanatto2025spatial}, it introduces ergonomic issues, limited input precision, and spatial overload~\cite{wang2025handows, ens2014personal,ens2014ethereal, pavanatto2025spatial}. 

Existing methods, including world-fixed~\cite{feiner1993windows, lee2013spacetop, 10.1145/3290605.3300577}, head-fixed~\cite{ens2014ethereal, ren2020understanding, lee2017projective}, and body-anchored layouts~\cite{wang2025handows, li2021armstrong, serrano2014exploring, yan2018eyes}, provide partial solutions but remain limited in adaptability. Many have added spatial layouts as a common way for window management, such as using 3D windows anchored to the world~\cite{lee2017projective, ens2015spatial}. However, spatial layouts can overload users when too many visual windows exist in their workspace, overwhelming attention with virtual window occluding each other and harming task efficiency~\cite{lee2013spacetop, ens2014personal}. SpaceTop~\cite{lee2013spacetop} tackles occlusion and z-order confusion by introducing a see-through desktop that distributes windows in depth above the keyboard, blending 2D and 3D interactions. Depth placement avoids overlap and supports a focus-plus-context workflow. Personal Cockpit~\cite{ens2014personal} instead reduces visual clutter by organizing windows on a body-centric ``cockpit'' scaffold around the user, using an egocentric layout that uses spatial memory and small head or body movements for quick access and task switching. Together, these systems show how immersive, spatially organized workspaces can balance flexibility, structure, and cognitive efficiency~\cite{pavanatto2024multiple}.

Another key challenge is input precision. As users interact across multiple spatially distributed windows, traditional pointing and tapping methods often become imprecise or inefficient~\cite{wolf2020understanding, hayatpur2019plane}. To address this, gaze-assisted systems, like Spatial Bar, combine eye tracking with cursor teleportation to enable rapid, precise switching between distant interface elements, reducing the need for large physical movements and improving selection accuracy~\cite{pavanatto2025spatial, ens2015spatial}. Complementarily, Handows enhances precision through embodied interaction. It anchors miniature window interfaces to the user’s palm, allowing stable, touch-based manipulation within a consistent, body-relative reference frame~\cite{wang2025handows}. Other approaches use spatial grouping metaphors that cluster nearby applications~\cite{robertson2004scalable}. Despite the effort, interacting with virtual windows still requires ergonomically personalized designs to support sustained, natural interaction in extended reality~\cite{10.1145/3586183.3606717}.

Our work builds upon the idea of Personal cockpit~\cite{ens2014personal} to use spatial anchors for swift virtual window placing, resizing and grouping, with the goal of reducing the interaction effort and improving efficacy.

\subsection{Workspace productivity with XR}
XR productivity refers to how immersive technologies are used to enhance knowledge work and digital collaboration by expanding the boundaries of traditional computing~\cite{luo2025documents,cheng2025augmented, biener2025long}. Prior research has explored this in mainly four domains. First, remote collaboration studies examined how immersive platforms like Horizon Workrooms support team communication compared to tools such as Zoom, highlighting deeper engagement but with issues of discomfort and sub-optimal usability~\cite{abramczuk2023meet, gauglitz2014world, fakourfar2016stabilized,teo2019mixed}; similar explorations extended to domain-specific medical teamwork~\cite{sadeghi2021remote, gasques2021artemis}. Second, long-term studies assessed sustained VR work over full weeks, showing potential for deep focus and distraction mitigation. However, challenges like simulator sickness~\cite{biener2022quantifying} hinders the practical use. Recently, new devices like Sightful’s Spacetop or Apple’s Vision Pro integrate XR workspace into everyday workflows without noticeable simulator sickness~\cite{cheng2025augmented, biener2025long}. Third, interaction research examined hybrid input methods such as adapting 2D mice for 3D tasks, combining gaze with touch on tablets, and exploring how virtual documents can be efficiently arranged and manipulated in depth~\cite{zhou2022depth, luo2025documents, biener2020breaking}. Finally, work on Extended Reality Visual Guidance explored training and industrial use cases by measuring how visual overlays influence task accuracy and cognitive load~\cite{pietschmann2025enhancing}.

The benefits and challenges of XR productivity reflect the trade-offs between immersive potential and current technological limitations. Key benefits include expanded workspace and visualization~\cite{ball2005effects,green2009building}. XR offers virtually infinite, customizable display space unconstrained by physical monitors~\cite{pavanatto2021we, biener2022quantifying}. It also enhances focus and efficiency by blocking real-world distractions, fostering deep engagement and flow~\cite{ruvimova2020transport, biener2025long}. In collaborative contexts, XR fosters a strong sense of social presence and immersion during meetings, improving connection and engagement over traditional video calls~\cite{abramczuk2023meet}. 

XR for productivity also entails portability and privacy, enabling users to work in almost any setting while maintaining confidentiality of sensitive content~\cite{zhou2022depth, pavanatto2021we, biener2020breaking, vijay2023hyway}.
However, several challenges persist. From an ergonomic and health standpoint, extended headset use causes discomfort, eye strain, and simulator sickness~\cite{pavanatto2021we, biener2022quantifying, liu2024using, biener2025long}. Technical limitations such as low resolution, narrow field of view, latency, and tracking errors undermine usability~\cite{pavanatto2021we, liu2024using, abramczuk2023meet}. Interaction and input remain problematic—midair gestures lack precision, and physical interactions like note-taking are awkward~\cite{zhou2022depth, biener2022quantifying, ruvimova2020transport}. Workflow integration issues arise from the absence of native XR software and the reliance on mirrored desktop setups~\cite{biener2025long}.

Window management in XR can introduce strain due to frequent head movements during multitasking~\cite{abramczuk2023meet}. Inconsistent 2D/3D input mappings and limited contextual control further reduce efficiency~\cite{zhou2022depth}, making window adjustment difficult. Inefficient window management harms productivity because it determines how efficiently users can access, compare, and manipulate information. Research shows that larger or multiple displays significantly enhance performance, especially for cross-referencing or transferring content between windows~\cite{pavanatto2021we}. Using virtual layouts can increase efficiency by reducing application switching time and allowing multiple tasks to remain visible at once~\cite{biener2022quantifying, biener2025long}. Addressing these gaps requires innovations in multimodal input design, ergonomic adaptation, and intelligent, context-aware spatial window management to realize XR’s full productivity potential~\cite{pavanatto2021we}.

\subsection{Rule-based constraints for UI adaptation}
A substantial body of work formulates MR/AR interface adaptation as decision-making under constraints to reduce cognitive load and improve XR work efficiency~\cite{10.1145/3332165.3347933, 10.1145/3332165.3347945, 10.1145/3586183.3606717, 10.1145/3290605.3300577, 10.1145/3491102.3517723}. These systems employ rule-based reasoning to modulate visibility~\cite{10.1145/3332165.3347945,10.1145/3544549.3585732, 10.1145/3746059.3747645}. Toolkits generalize these ideas: AUIT lets experts encode objectives and resolve conflicts qualitatively during adaptation design~\cite{evangelista2022auit}, XRgonomics optimizes placements for comfort in 3D UIs~\cite{10.1145/3411764.3445349}, and RealityCheck blends virtual and real contexts by compositing real-time 3D reconstructions into VR~\cite{10.1145/3290605.3300577}. Common across this literature is a formalization of UI adaptation as an optimization problem with explicit objectives and constraints (e.g., visibility vs. occlusion, salience vs. distraction, comfort vs. reach), enabling principled trade-offs~\cite{10.1145/3586183.3606717, lindlbauer2019context, cheng2021semanticadapt}. Collectively, these approaches advance the field by reducing attentional fragmentation and interaction cost, improving task efficiency, and providing designers with scalable methods that move beyond ad-hoc heuristics~\cite{fender2018optispace,fender2017heatspace, gajos2005fast}.

We optimize window management by taking users' physical orientation and interaction history to auto-adjust the layout of virtual windows, reducing the need for participants to perform the tedious work of window resizing. 

\subsection{LLMs for interface planning, recommendation, and explainability}
Recommendation for adaptive and generative interfaces in XR remains challenging in deciding what content to surface, where and how to place it, and when to adapt\cite{cho2023realityreplay}. Recent systems leverage LLMs to move from descriptive prompts to prescriptive layouts and executable code, such as LayoutGPT~\cite{layoutgpt}, Layout Prompter~\cite{lin2023layoutprompterawakendesignability}, UI Grammar~\cite{lu2023uilayoutgenerationllms} and UICoder~\cite{wu2024uicoderfinetuninglargelanguage}. This allows users to be more efficient without the need to describe everything in text. 

In XR spaces, LLMs can also reason about situated context to provide adaptive and contextually appropriate UI behaviors. They dramatically lower cognitive load, cost for semantic understanding, and enable end‑to‑end pipelines. SituationAdapt~\cite{10.1145/3654777.3676470} introduces a VLM‑based reasoning module that judges occlusion, social appropriateness, and safety while AgentAR~\cite{10.1145/3746059.3747676} uses LLM-based autonomous agents that incorporate user's high-level goals, context, and interaction history to dynamically generate and adapt AR action step-guidance. While focused on UI generation, Chen et al.~\cite{chen2025generativeinterfaceslanguagemodels} relies on LLMs to translate user queries, which inherently contain task context, into adaptive and interactive UIs. Our work also uses LLMs for processing context information to generate adaptive workspaces.


\section{User agency in human-AI collaboration}
In the context of human-AI collaboration, User Agency refers to a person’s perceived ability to influence and control an interactive system, especially when collaborating with autonomous or AI-driven components~\cite{zhou2024coplayingvr, zhang2024vrcopilot, zhou2025juggling}. It represents a balance between human intention and system automation, shaped by perceptions of control, embodiment, and input strategy. Perceived control defines how consciously users feel in command of their actions and environment~\cite{zhang2024vrcopilot}. Research shows that direct, manual control fosters stronger agency and embodiment, particularly in critical tasks, while high automation can reduce ownership and engagement~\cite{zhou2025juggling}. However, scaffolded approaches---where AI presents intermediate representations like wireframes---can enhance perceived agency by keeping users in the creative loop~\cite{zhang2024vrcopilot}.


Meanwhile, user agency is deeply linked to how windows should be managed, as the agency determines how users perceive control, responsiveness, and fluency in commanding these windows~\cite{10.1145/3746059.3747676, ens2014ethereal, schmalstieg2000bridging}. Effective window management requires dynamic, context-aware adaptation that minimizes friction between manual and automated actions~\cite{10.1145/3586183.3606717, pei2024ui, lystbaek2024spatial}. Systems must continuously adjust interface elements to suit the user’s context, task demands, and engagement levels. For example, proactive AR agents need to shift their communication mode from visual icons to verbal confirmations or contextual prompts based on urgency or attention~\cite{10.1145/3746059.3747748, wang2025handows, lystbaek2024spatial}. Poorly aligned UIs, such as menus outside the user’s natural gaze or overlapping objects, can disrupt ergonomics and weaken agency with unnecessary users efforts and attention shifts~\cite{pavanatto2021we, robertson2005large, ens2014personal}. Managing multiple workspaces further reinforces agency by enabling flexible control over complex virtual scenes. Techniques like multi-workspace visualization allow users to switch between multiple design states seamlessly. For example, World in Miniature provides an overview that enhances spatial awareness and control without breaking immersion~\cite{zhang2024vrcopilot}. Similarly, embodied authoring through interactive design enables users to shape and adjust elements in real time, sustaining a sense of immediacy and creative ownership\cite{lee2017projective, schmalstieg2000bridging, ens2014ethereal}. Together, these approaches show that maintaining user agency in mixed-initiative and immersive systems depends on adaptive, ergonomic, and embodied window management that preserves user control while leveraging automation for fluid interaction. Our implementation offers users full agency while enabling automation via natural language to convey the high-level goals, and simple interaction to obtain the AI's recommendations.

%% file: sections/03_system.tex
\begin{figure}[t]
    \centering
    \includegraphics[width=1.0\linewidth]{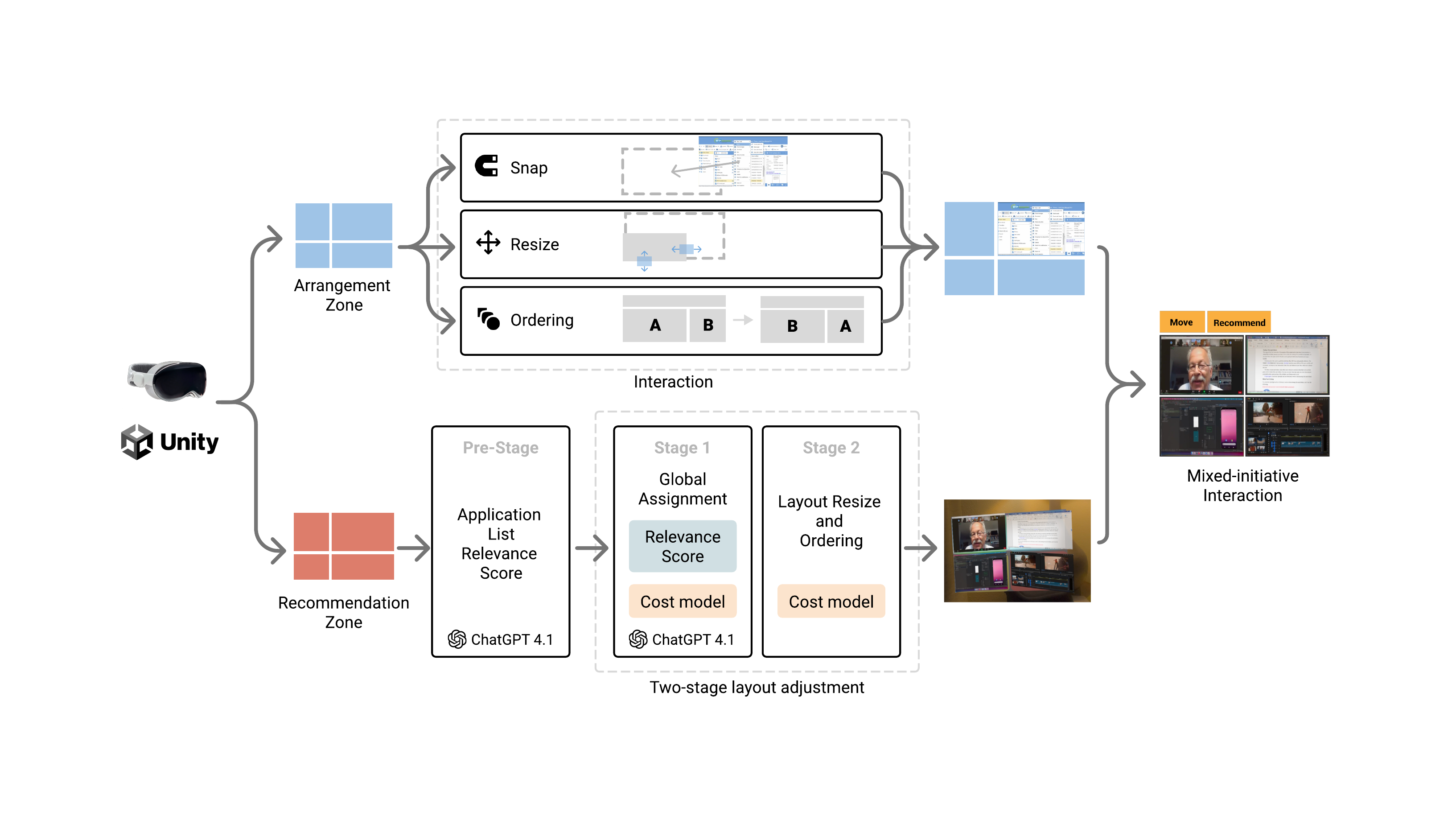}
    \caption{System flow chart to indicate how different spatial configuration contributes to a mixed-initiative experience.}
    \label{fig:system}
        \vspace{-8pt}
\end{figure}
\section{Design Goals}




\subsection{Enabling speed, anchoring, layout, and reducing effort}
\textbf{Using spatial anchors to help improve speed.} Enabling users to statically arrange their XR workspace is crucial for productivity, especially since manipulating virtual windows in 3D can be slower than classic 2D interactions~\cite{pavanatto2025spatial}. Rapid layout authoring in XR is difficult due to spatial complexity, depth perception, and limited input precision, even for simple interactions such as moving virtual windows with hand gestures~\cite{serrano2014exploring}. Lacking an obvious anchoring point for spawning and positioning in an XR environment is one of the core reasons behind the interaction challenge. 

\textbf{Enabling diverse layouts for spatial organizations.} The purpose of having layout is to facilitate complex tasks and multitasking. The layout should be easy to create, adjust and position in 3D space, with minimal time spent to manage virtual windows. In addition, it should offer functions to help users group windows to reduce disorientation~\cite{robertson2000task, ens2014ethereal}. Layouts should also be able to support variable application types, allowing apps dominantly vertical or horizontal to reside comfortably within. Finally, interacting with layout should be intuitive and easy to use. 

\textbf{Reduce effort improves efficiency. } Users often need to adjust the virtual window's size, position to fit the progress of the work. These adjustments are typically done with mid-air pointing (i.e., ray casting~\cite{cho2023realityreplay}) and hand gesture, and sometimes keyboard and mouse are also used for more precise input. However, the large canvas in XR still poses more physical demand on all types of input. Being able to reduce the effort is key to bringing the 3D window manipulation experience closer to its 2D counterpart. 

\textbf{Reducing cognitive load is critical for productivity. }  Multitasking within virtual environments escalates the user's mental workload~\cite{setu2025predicting}. The complex interaction required—combining physical navigation, 3D manipulation, and simultaneous cognitive tasks—is highly taxing on a user's attention, working memory, and physical capacity~\cite{juliano2022increased}. One goal is to minimize the physical effort by avoiding large arm movements~\cite{wang2025handows}. Meanwhile, simplifying interaction and reduce the number of repeated interactions for window adjustment can help reduce the cognitive load. Moreover, optimizing visual information delivery should help users to focus on critical information, resulting in lower effort to identify useful data. 

\subsection{User-centered human-AI collaboration}
\label{human-ai collaboration design}
\textbf{Improve efficiency while keeping users in control.} The goal is to use AI for user productivity and experience while avoiding overwhelming them with digital clutter or unpredictable automation. As such, we adopt a mixed-initiative strategy~\cite{teo2019mixed}. Most prior work in automated window management gives g users little room to control the window layouts. For window layout management, it is important to let users in control of the overall layout because in a work situation window movement without users' intention causes confusion and requires extra work for the user to update their mental state. Meanwhile, users should easily command AI for efficiency related tasks such as app selection or layout adjustment. But users should also easily override AI's recommendations to retain agency~\cite{caetano2025design}.

\textbf{Context is important.} XR environments are dynamic; AI needs to adapt to users' context, such as user's physical location, orientation, and interaction state. This helps AI to provide the right information in the right time~\cite{davari2024towards}.This would allow the system to automatically adjust the layout, size and content of the window and help users reduce window management workload. 

\textbf{User's intent can be useful.} Inferring intent is inherently uncertain. One aspect of the goal is to align the automated results with the user's interaction process. For example, when the user is trying to establish a new workspace, the AI should help users achieve that goal faster with ample room for user fine-tuning. While salient information like gaze may indicate interest, but they can be ambiguous. It is important to obtain detailed intent from user for the system to automatically recommend useful applications~\cite{10.1145/3544548.3580873}.  Using LLMs helps to understand the semantic meaning from the users' intent and reason among different applications to fit the intent~\cite{10.1145/3654777.3676470}. 


\subsection{Dynamically readability suggestions.}
\label{DG3:readability}
Preserving readability of virtual content is crucial in window adjustment and arrangement for users to digest information. As a larger window and content size leads to better readability in XR, it also results in more effort to navigate, larger degrees of head and hand movement, resulting in higher interaction cost. Font sizes should be carefully considered~\cite{abramczuk2023meet} to ensure minimum readability as well as the overall size of the virtual windows. In addition, content size and background color, field of view (FOV), and lighting can all affect how users read the content.

In addition, the large canvas in XR is considered for readability since there is a trade-off between interaction cost (e.g., head and hand movement) and display size. As XR canvas is intrinsically 3D, virtual content closer to the user becomes larger and more readable, but cost more to navigate. As a result, semantically relevant windows (e.g. browser to notes, IDE to terminal) should be placed close to each other for readability. 

\subsection{Occlusion-free area.}
The goal of this area is to let users create a region for direct see-through, allowing them to receive information from physical environment. Users should be given chance to avoid virtual interface occlusion. This has been found critical in prior work~\cite{10.1145/3654777.3676470,davari2020occlusion}. The area should be easily configurable into different sizes to fit user needs. Also the occlusion-free area should provide proper warning when interfaces ``intrude''.

\begin{figure}[h]
    \centering
    \includegraphics[width=0.9\linewidth]{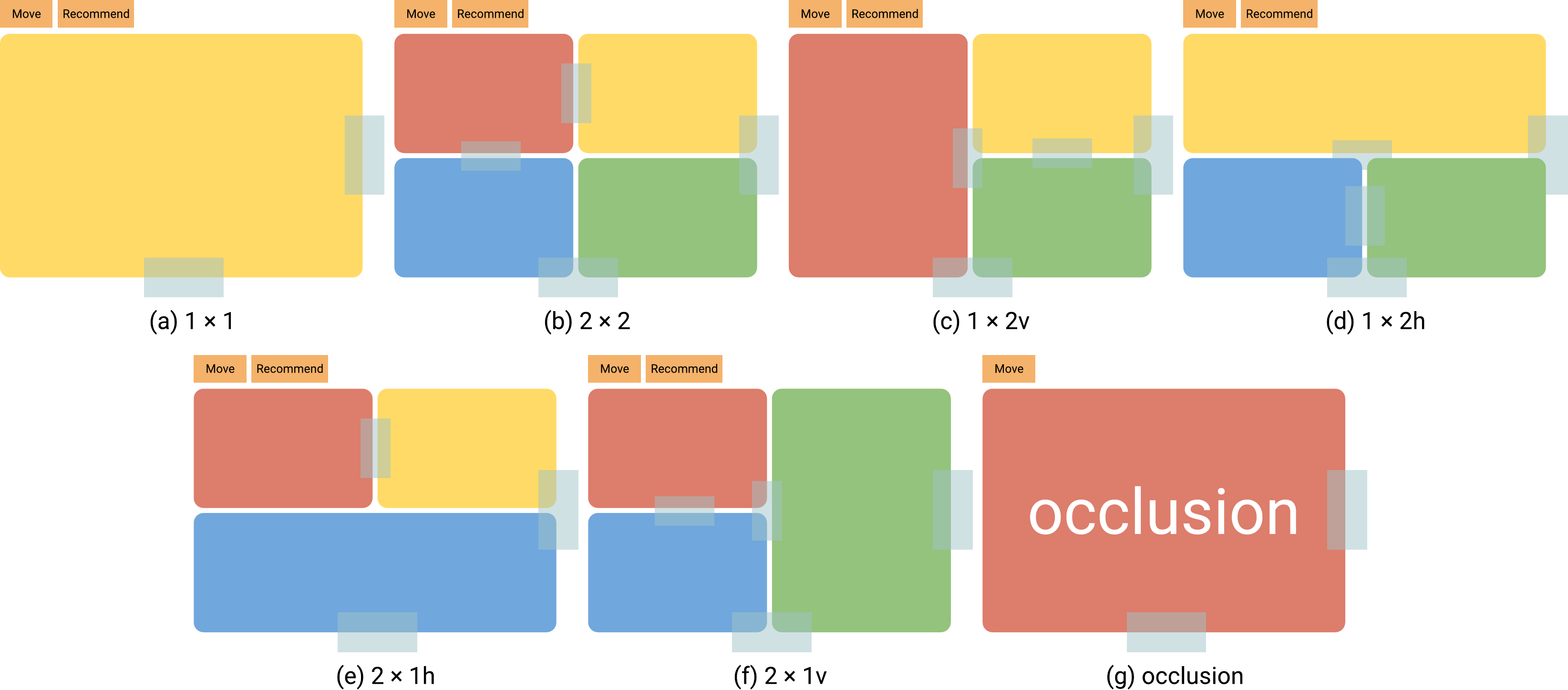}
    \caption{Extending the tiled window convention, we use six different layout templates as 3D spatial anchors to support window resize, ordering, snapping, and grouping. Subfigures a) to f) indicate different tiling. For all these templates, the left upper corner tile is considered w0 and h0, with the width of the template defined as W and height as H. It is important to note that the same row cells share the same height, and the same column cells share the same width. g) An occlusion template placed in XR space prevents other zone templates from entering, allowing users to see through without virtual information blocking the view. }
    \label{fig:layout}
        \vspace{-8pt}
\end{figure}




\section{DuoZone System}
Instead of providing a fixed window management plan for XR, DuoZone aims to achieve efficient window management through customized and human-AI collaborative spatial configurations. We extend upon the idea of spatial configurations~\cite{ens2014personal} to allow users to customize for different layouts, accommodating for types of work (e.g., design, engineering, document organization).

\begin{figure}[htbp]
    \centering
    \includegraphics[width=1.0\textwidth]{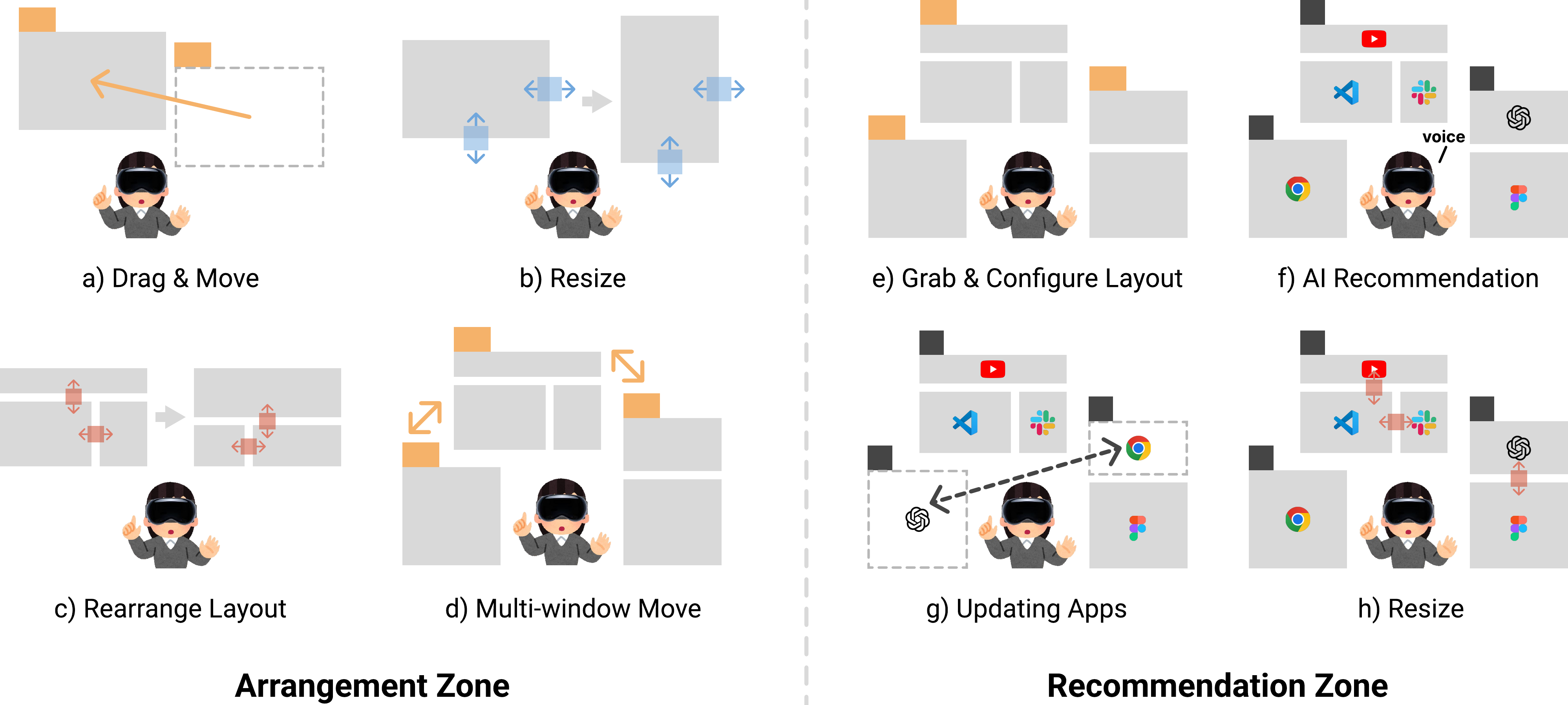}
    \caption{Interactions in DuoZone’s Two-Zone Framework. In the Arrangement Zone (a–d), users directly drag, resize, rearrange, and move multiple windows to create personalized layouts. In the Recommendation Zone (e–h), users can voice requests for AI-generated layouts. The AI will automatically recommend apps, and fine-tune window sizes. Together, these modes balance user control with AI assistance for efficient workspace organization.}
    \label{fig:interaction} 
\end{figure}

We use zones to reference spatial configurations; and zones are intractive layout templates enabling resizing, scaling, and contextual auto-filing XR applications in designated regions (see figure~\ref{fig:interaction}). These zones offer functions beyond pre-set rules, making users creators rather than operators of the XR workspace. 


\textbf{Section overview.} We will first present the design and implementation of two zones, \textit{Arrangement Zone} and \textit{Recommendation Zone}. In recommendation zones, we describe a multi-stage LLM-based cost model for application recommendation, and the zone's automatic sizing and layout adjustment. Afterward, subsections will describe the system setup and data flow. 

\subsection{Arrangement Zone}
As a foundation to support configurable workspace, arrangement zones are translucent, window-like spatial layouts in XR. These layouts function as adjustable containers allowing users to manually adjust their inner cells, overall size, and positions. Inspired by the design of tiled windows~\cite{bly1986comparison}, we designed six layout templates that are commonly used in modern operation systems to distribute space within a zone. These six templates covers basic horizontal and vertical tiling windows and is not inclusive. In practice, users can construct more complex layout from these templates and assemble into more templates based on the needs. Spatial areas covered by arrangement zones will provide semi-automatic organization; other spaces in XR will remain fully manual control by users. This setup allows users to actively plan for their workspace layout with their intended needs and functions. 

\subsubsection{Interaction.}
Interacting with the arrangement zone follows the traditional drag-and-drop metaphor. This interaction language is consistent with how users move windows or applications on a desktop, allowing for a continuous mental model when switching working conditions from desktop to XR. In addition, the arrangement zone supports a set of interactions commonly used (e.g., snapping, alignment, grouping), aiming to improve the efficacy of common micro window operations (DG1). Below are lists of supported interactions:
\begin{enumerate}
    \item \textbf{Dragging a virtual window into} a cell of a zone adds this window into the zone, resulting in automatic snapping, resizing, and re-orientation of the window. 
    \item \textbf{Dragging a virtual window out} from a cell decouples them, with the window's size and orientation inherited from previous state.
    \item \textbf{Moving inner knobs} between cells redistributes their sizes. For example, in 3 to 7 ratio grid of two cells, moving the inner knobs like a slider will change their ratios, hence the size of the two cells. 
    \item \textbf{Moving outer knobs} allows for resizing the zone both horizontally and vertically, virtual windows inside the zone will resize proportionally.
    \item \textbf{Translation} is supported via drag-and-drop a \textit{Move} button on the left-top corner of the zone, allowing both the zone and its containing app to move as a group.
    \item \textbf{Rotation} is automatically calculated using Unity's \textit{LookAt} function. This allows us to achieve efficient XR workspace interactions similar to \textit{PersonalCockpit}~\cite{ens2014personal} where zones always face the user.
\end{enumerate}

These interactions extend Apple's PolySpatial library in Unity. We use the library's native events that capture ray-casting results from gaze and gesture world coordinates or mouse pointer locations. These coordinates are used to initiate interface state changes (e.g., drag start, drag end, taps/clicks, hover) of the virtual window and arrangement zones. For example, once the user drags a window while hovering on a cell, it highlights to indicate a change of state, indicating a snapping follows.

\subsubsection{Occlusion-free template}
We designed a type of arrangement zone that allows users to label a ``void'' area in XR. This area will be directly see-through, allowing users to receive visual information from the environment directly. Similar to a 1x1 layout, the user can move, adjust the size, and rotate the occlusion-free zone upon placement. This zone has a red-filling color and will become opaque once the user is not interacting with it. When other zones or virtual windows intersect with this zone, a red contour encompassing the zone will be visible, and dropping a window or zone at this time will move them to the side. A user can place as many occlusion-free zones as they want without overlapping. 

\subsection{Recommendation Zone}
Aside from aiming to improve manual window management, we designed another type of zone focus on providing contextual automation. This gives user an alternative mentality to ``outsource'' some of the management burden. Meantime, the having zones dedicated for recommendation allows users to keep autonomy and decide what not to be automated, reducing the risk of loss of agency during an human-AI collaboration (See figure \ref{fig:system}).  

The core challenge of the AI recommendation has two parts: 1) to decide what virtual window app to put in which cell of multiple zones, and 2) how to determine the size, order, and layout of zones. We do not want to change the location of the recommendation zones to retain the users' spatial mental model~\cite{cherep2020spatial}, but allowing for size changes to optimize potential interaction cost in XR environment. Following subsections will describe a two stage interaction model. 

We begin with users created $K$ recommendation zones in XR; each zone $Z_k$ ($k = 1, \ldots, K$) is characterized by:
\begin{itemize}
\item \textbf{Layout} $\tau_k \in \mathcal{T}$, where $\mathcal{T} = \{\text{1×1}, \text{2×2}, \text{1×2v}, \text{1×2h}, \text{2×1v}, \text{2×1h}\}$, see figure~\ref{fig:layout} for details.
\item \textbf{Dimensions} $(W_k, H_k)$ as width and height in meters
\item \textbf{3D transform} $(p_k, o_k)$ where $p_k \in \mathbb{R}^3$ is position and $o_k \in SO(3)$ is orientation relative to the user
\item \textbf{Cell set} $\mathcal{C}_k = \{c_{k,1}, c_{k,2}, \ldots, c_{k,n_k}\}$ where $n_k$ is the number of cells determined by $\tau_k$
\end{itemize}

\subsubsection{pre-Stage: LLM-based Application Relevance Prediction}
Our system uses ChatGPT 4.1 for semantic and goal-aware reasoning from user's high-level goal. When the user triggers an AI recommendation, we use Meta's Wit for voice-to-text and provides LLM with the transcribed text. Users are given a chance to edit the text via our UI interface if needed. Based on this goal ($G$), LLM will output a set of recommended virtual window applications $\mathcal{A} = \{a_1, a_2, \ldots, a_N\}$. Meanwhile, an accompanying set of relevance score $\mathbf{r} = \{r_1, r_2, \ldots, r_N\}$ where $r_i \in [0,1]$ represents the predicted likelihood that application $a_i$ will be used. 

\subsubsection{Constraints}
We define the following constraints for further two stage interaction models. For sizing constraints, it is important to note that: 1) Each cell contains at most one application, but could be empty; 2) The number of applications assigned to zone $k$ cannot exceed its cell count; and 3) layout's dimension must within the zone dimensions. For readability, each cell must maintain a minimum angular size to ensure text readability. Following Microsoft's MR Typography guidance for readability, we require that the smallest font-size on applications to be more than $0.5$ degrees (Section \ref{DG3:readability})  from users' point of view~\ref{eq:readability}. 

\begin{equation}
\label{eq:readability}
\min\left(\arctan\left(\frac{w}{d}\right), \arctan\left(\frac{h}{d}\right)\right) \geq \alpha_{\min}
\end{equation}

\subsubsection{Interaction Cost Model}
We implemented a two-stage cost model to allow easy navigation among the applications, decide on the adjacency, and determine functional size and ratio-configuration of an application (e.g., an IDE would likely to be placed in a landscape manner and be larger than others). These automations are important for XR usability and less effort to set up the workspace, while semantic application selection and functional size ensure minimal user adjustment for the workspace (see design consideration in Section~\ref{human-ai collaboration design}).~First, we define four cost signals that capture different aspects of interaction effort:

\begin{enumerate}
    \item \textbf{$F_{ij}$: Pointing Distance.} The Euclidean distance between the centers of cells containing applications $a_i$ and $a_j$. For cells in the same zone $k$ with centers at positions $\mathbf{c}_i^k$ and $\mathbf{c}_j^k$:
        \begin{equation}
        F_{ij}^k = \|\mathbf{c}_i^k - \mathbf{c}_j^k\|
        \end{equation}
        
        For cells in different zones, we use the distance between zone centers $\mathbf{p}_k$ and $\mathbf{p}_\ell$:
        \begin{equation}
        F_{ij}^{\text{zone}} = \|\mathbf{p}_k - \mathbf{p}_\ell\|
        \end{equation}
        
        This distance serves as a proxy for pointing and gestural input costs, as greater distances require larger movements and typically longer selection times~\cite{McCracken:1990:SSC:575315}.

    \item \textbf{$H_j$: Head Turn Angle.} The angular displacement required to bring the user's gaze from the current forward direction to the center of the cell containing application $a_j$. Let $\mathbf{v}_{\text{forward}}$ be the user's forward viewing vector and $\mathbf{v}_j$ be the direction vector from the user to cell $j$:
    \begin{equation}
    H_j = \arccos\left(\frac{\mathbf{v}_{\text{forward}} \cdot \mathbf{v}_j}{\|\mathbf{v}_{\text{forward}}\| \|\mathbf{v}_j\|}\right)
    \end{equation}

    \item \textbf{$M_j$: Hand Movement Distance.} The Euclidean distance from the user's current hand position $\mathbf{c}_{\text{current}}$ to previous hand position  $\mathbf{c}_{\text{prev}}$ between a \textit{pointerdown} and \textit{pointerup} event. Since Vision Pro does not require users to lift their hand to perform linear direct manipulation, hand movement could be non-linear when comparing to distances between zones.
\end{enumerate}

 In practice, we use a list of dictionary to store these cost signals in a time sequence. The list also contains a normalized $\widetilde{F}_{ij},\widetilde{H}_j, $ and $\widetilde{M}_j $ that are determined from the zone's layout and configuration. 

 \textbf{Stage 1.} We use the reasoning capabilities of ChatGPT-4.1 to recommend a global application assignment across all cells and zones. Rather than following an algorithmic approach, this approach considers nuanced trade-offs, contextual relationship between applications, and users potential cost that could be difficult to encode in optimizing objectives. We first compute a cost matrix using initial layout parameters $\boldsymbol{\Theta}^{\text{init}}$ (layouts with default splits or sizes given by the user):
 
\begin{equation}
C_{i,k,j} = \sum_{\ell \in \mathcal{A}_{\text{prev}}} \left[ r_i r_\ell P_{i\ell} \cdot c_{i \to \ell}^{k,j} +  r_\ell r_i P_{\ell i} \cdot c_{\ell \to i}^{k,j} \right]
\end{equation}
 
Where $r_i r_\ell$ denotes the joint relevance of applications (obtained from previous LLM call) $i$ and $\ell$, representing the likelihood that both applications will be used together. Furthermore, $P_{i\ell}$ is the transition frequency from application $i$ to $\ell$ and $c_{i \to \ell}^{k,j}$ is the realtime transition cost from application $i$ (hypothetically placed in cell $(k,j)$) to application $\ell$. Finally, $\mathcal{A}_{\text{prev}}$ is a set of previously assigned applications.

The instantaneous cost is computed as following, with all cost weights using equal weights from pilot experiments:
\begin{equation}
c_{i \to j}^{k,\ell} = 
\begin{cases}
\lambda_f \widetilde{F}_{ij}^k + \lambda_h \widetilde{H}_j^k + \lambda_m \widetilde{M}_j^k  & \text{same zone } k \\
\lambda_f \widetilde{F}_{ij}^{\text{zone}} + \lambda_h \widetilde{H}_j^{\text{zone}} + \lambda_m \widetilde{M}_j^{\text{zone}}  & \text{different zones}
\end{cases}
\end{equation}

\textit{LLM Prompting Decisions.} Along with the cost matrix described above, we construct the prompt with the following data: 1) recommended application list with relevance scores from the first LLM call; 2) user created zones with their layout, sizes, and all occupied cells; 3) the cost matrix $C_{i,k,j}$; 4) readability constraints; 5) high-level goal $G$ for context. The LLM's output consider multiple factors simultaneously, such as relationship and appropriateness along with numerical data. For example, LLM might agree with that ``emails and calendars should be placed adjacent'' even if the pure cost matrix suggest otherwise. 

\textbf{Stage 2.} Our preliminary testing found that using only prompt from LLMs yield random results in resizing the zone's layout or size, even if we provide information for users' distances or angular distances. As a result, after we obtain the assignment results from stage 1, each zone $k$ independently optimizes its internal layout parameters. Let $\mathcal{A}_k = \{a_i : \sigma^*(a_i) \in \mathcal{C}_k\}$ denote applications assigned to zone $k$. The goal here is to find a layout configuration $\boldsymbol{\theta}$ where:
\begin{equation}
\label{eq:stage2}
\boldsymbol{\theta}_k^* = \arg\min_{\boldsymbol{\theta}_k \in \Omega_{\tau_k}} \left[ \mathcal{C}_k^{\text{local}}(\boldsymbol{\theta}_k \mid \sigma^*) + \lambda_s \mathcal{S}_k^{\text{local}}(\boldsymbol{\theta}_k \mid \sigma^*) \right]
\end{equation}

Here $\boldsymbol{\theta}$  represents the point $(w_0, h_0)$ within a zone where horizontal divider crosses $w_0$ and vertical divider crosses $h_0$. Each zone is resized locally based on its configuration and layout. To do that, we follow:
\begin{equation}
\mathcal{C}_k^{\text{local}}(\boldsymbol{\theta}_k \mid \sigma^*) = \sum_{i,j \in \mathcal{A}_k} r_i r_j P_{ij} \cdot c_{i \to j}^{\text{intra}}(\boldsymbol{\theta}_k)
\end{equation}

where:
\begin{equation}
c_{i \to j}^{\text{intra}}(\boldsymbol{\theta}_k) = \lambda_f \widetilde{F}_{ij}^k(\boldsymbol{\theta}_k) + \lambda_h \widetilde{H}_j^k + \lambda_m \widetilde{M}_j^k 
\end{equation}

We then calculate the local size-relevant term, which encourages allocating more space to higher-relevance applications within zone $k$.
\begin{equation}
\mathcal{S}_k^{\text{local}}(\boldsymbol{\theta}_k \mid \sigma^*) = -\sum_{i \in \mathcal{A}_k} r_i \cdot A_i^k(\boldsymbol{\theta}_k)
\end{equation}

A final step after finding the results of the most optimal $\boldsymbol{\theta}_k$ is to conditionally scale-up the entire zone if the smallest cell in the zone does not match the readability constraint (Equation: ~\ref{eq:readability}). Finally, the layout can be updated using $W_0$ and $h_0$ to the following, see figure~\ref{fig:layout} for visualization:
\begin{equation}
\begin{aligned}
\text{Cell } P_0 &: w_0 \times h_0 \\
\text{Cell } P_1 &: (W - w_0) \times h_0 \\
\text{Cell } P_2 &: (W - w_0) \times (H - h_0) \\
\text{Cell } P_3 &: w_0 \times (H - h_0)
\end{aligned}
\end{equation}

\subsubsection{Front-back Communication}
The Vision Pro runs a Unity instance that sends the \textit{interaction logs} and \textit{the high-level intention} to a python server in JSON format. The python server then communicates with ChatGPT 4.1 using its native API. On average, each request takes about 4 seconds to receive a feedback on a WiFi network. The resulting information is sent from the python server to Vision Pro's Unity instance via http request.

\subsubsection{Interaction.}
Interacting with the recommendation zones require users to convey a \textit{high-level} goal via voice or text input, such as ``coding a project'' or ``designing a website in a multi-user project''. This ensures that the backend AI automatically set ups the workspace based on users' needs. Once the backend AI recommend virtual windows for the user, we added a confirmation stage for users to decide whether to accept the recommendations. This design aims to provide a non-intrusive experience with AI's recommendation temporarily occupies the zones but users keeping the full agency.

%% file: sections/04_user_study.tex
\section{User Study}
To understand DuoZone's efficacy and its mixed-initiative design's potential for XR window management, we conduct an empirical study with expert users who worked under multi-window situations in their daily lives. We use a within-subject design and devised two primary tasks. These tasks elicit users' performance and the cognitive and physical effort of our system. We ask the following research questions:


\begin{enumerate}
    \item Whether DuoZone improves the interaction speed and users' cognitive load?
    \item Whether DuoZone reduces the number of manual adjustments to virtual windows
    \item In what ways DuoZone's recommendation can be useful?
    \item How does DuoZone's recommendation compare to users' manual setup?
\end{enumerate}

\subsection{Apparatus} 
We used an Apple Vision Pro (1st-generation) as the XR headset that connects to our \textit{DuoZone} client. The client was implemented in Unity using Apple's PolySpatial integration so that gaze and gesture events from visionOS could be captured at world-coordinate precision for snapping, drag, and other window operations. A dedicated host equipped with an NVIDIA GeForce RTX~4050 GPU handled real-time AI assistance, cross-process communication, and threading. The headset and server communicated over the local network for low latency data transfering. 


\subsection{Participants}
We recruit 16 participants (8 males and 8 females) between 21 to 39 years old (M=26.7, SD=$5.5$). They all had at least three years of professional working experience and use multi-window configuration in their daily jobs. We use snowball sampling and send digital flyers on social media and direct messaging tools. All but one participant had prior experience with XR interaction, and experienced with virtual windows before this experiment. 

\subsection{Tasks}
\subsubsection{Task 1: Layout Matching}
Participants performed a precise layout matching task. There are 15 trials per condition (Baseline vs. DuoZone), for 30 trials total. In each trial, 3–4 application panels (e.g., browser, notes, chat) appeared in a compact, overlapping start state centered in front of the user. A holographic target indicator then specified a goal configuration: a target sector in space (one of Left, Left-Up, Up, Right-Up, Right), a layout family (e.g., 1×2, 2×1, 2×2), and per-panel target layout ratios.

Using both conditions, our user moved, aligned, and resized panels to match the target and then confirmed completion. The task isolates XR window-management micro-operations (drag, snap, resize, swap) to test whether zones reduce time and effort without confounds from higher-level planning; dependent measures included time-to-complete, number of edits, post-edit distance to target, readability compliance, and travel proxies.

\subsubsection{Task 2: Workspace Construction}
Each participant constructed a task-ready workspace for their assigned role as a programmer or designer under two conditions: a baseline Manual setup completed first and an AI-assisted setup using DuoZone to enable within-subject comparison. In the manual condition, users selected needed apps from a ~20-item palette, placed and sized apps entirely by hand until declaring ``ready''. In DuoZone condition, users created empty Zones, issued a high-level voice command (e.g., ``set up for coding a web game'') to receive LLM-proposed window-to-cell assignments, and adjusted suggestions before confirming readiness. This task examines whether DuoZone shortens time to ready, yields lower interaction cost layouts, and reduces mental demand while preserving agency, and also logs suggestion acceptance and subsequent edits.

\subsection{Conditions}
Based on the most recent window position, resizing and adjusting management interaction metaphor used in Meta Quest Pro, HoloLens 2, and Apple's Vision Pro, we used a manual adjustment as the \textbf{baseline} for the experiment. This baseline uses Vision OS's native interaction trigger (eye gaze and hand movement) and gives user full control of the virtual windows, similar to when they use other applications on Apple's Vision Pro. For DuoZone conditions, we use arrangement zones in Task 1 and recommendation zones in Task 2.

\textbf{Balancing and Order. } To reduce the learning and ordering effect, we counterbalanced the conditions with alternation for Task 1. Within Task 1's trials, we used a pre-generated table that randomized and balanced both the direction and layout in five directions, so that total number of layouts and directions tested are equivalent. There is no learning or ordering effect in Task 2. DuoZone condition is recommended by AI, and since we want to understand whether AI recommended useful things that are unexpected by the user, we always start with the manual condition. 

\subsection{Procedure}

All sessions were conducted in a quiet, well-lit indoor lab space with ample open area to permit natural head and hand movement. Participants performed the study tasks while wearing the Vision Pro, either in the \textit{DuoZone} setup or the \textit{Manual} baseline described above.

Participants who required vision correction used contact lenses or compatible optical inserts. The headset fit (strap tension and eye relief) was adjusted for comfort prior to task execution.

\subsection{Data Collection}
\textbf{Timed-performance: } We wrote a script in Unity that captures time difference between a pointer-down to pointer-up event. This difference captures when a drag event begins to a drag event ends, useful to detect the time it takes to resize a virtual window or zone layout, or to detect the drag and drop of a virtual window, etc. The script generates one entry per each action performed for an example of the log. For task 2, the time is calculated in total number of seconds to complete the setup. 

\textbf{Cognitive load: } We use a digitized NASA-TLX form that follows the original 0 to 100 TLX score. Following the original TLX, each of the six category questions is designed as a $21$ tick slider, with the minimum increment of $5$. TLX form is collected after each condition is performed in a task.

\begin{figure}[h]
    \centering
    \includegraphics[width=0.9\linewidth]{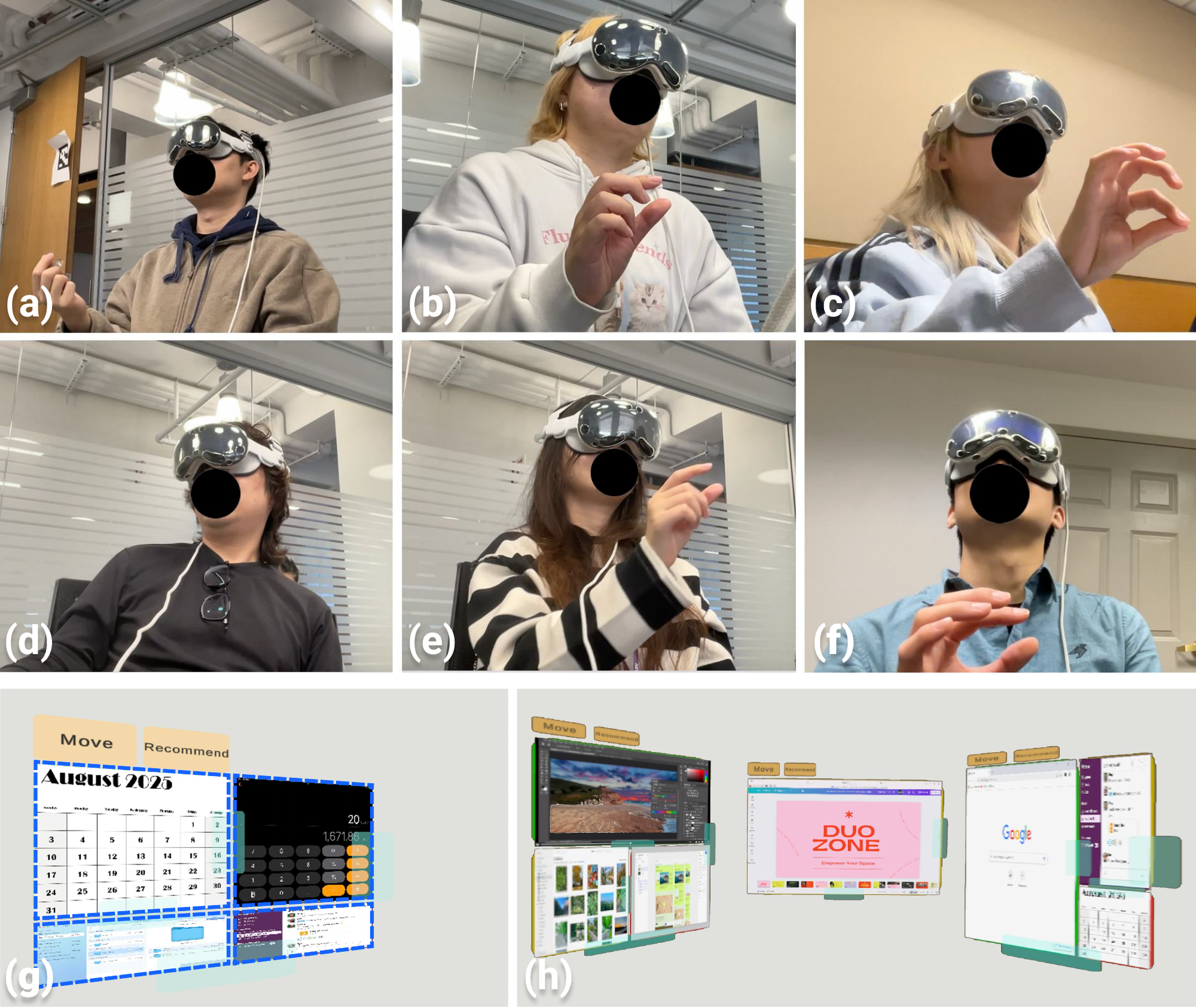}
    \caption{(a–f) Participants performing interaction tasks while wearing the mixed-reality headset. Each participant manipulates virtual windows using mid-air hand gestures to complete layout configuration tasks. (g) Task 1 interface, where participants were instructed to drag and place applications into designated target locations (top, bottom, left, right) to construct layouts with varying sizes. (h) Task 2 interface where participants configured either a design or programming workspace by arranging multiple applications into an integrated, ready-to-work environment.}
    \label{fig:user_study}
        \vspace{-8pt}
\end{figure}

\textbf{AI Recommendation Acceptance: }
We measure AI acceptance from recommended application acceptance, ordering and adjacency, and size and layouts. For acceptance, a script is used to record button responses from whether a user accept or decline application recommendations in each cell of a zone (and there is a button to batch accept per zone). The rest of the data are identified from the Vision Pro's first person recordings by two separate researchers. The final count is combined with any disagreement resolved through discussion. Researchers look for changes in user-performed re-ordering, changes in zone' size and inner layout proportion. Each user adjustment will be count as once. In every participant's session, two researchers report the total number of counts and total possible number of counts (i.e., if AI get it all wrong, how many actions users need to fix the situation). 

\textbf{Semi-structured interview: } We collect semi-structured interview questions at the end of each task. We recorded both the audio and video of each experimental session and use YouTube's auto-transcription function to extract transcription from the videos. 

In total, we collected 480 trials from 16 participants for Task 1. All participants completed two different conditions in Task 2, constitute 32 workspace setups. We have recorded a total of 32 videos (one third-person and one for first-person) with each about 1.5 hours. 

\section{Results}
\subsection{Timed-performance}
We log-transformed Task 1's completion time and a paired t–test revealed significant main effect of \textit{Condition} ($t(15)=6.22$, $p < 0.001$). On the original scale, the geometric–mean time ratio of DuoZone over Baseline was $0.655$ with a 95\% CI of [$0.566,\,0.757$], indicating that \textit{DuoZone} performs $34.5\%$ faster performance on average.

Within each \textit{condition}, we compared all pairs of orientations using two–sided paired $t$–tests on the log-transformed time; $p$–values were Bonferroni–adjusted within condition (10 comparisons per family). In \textit{Baseline}, several orientation pairs showed significant differences in performance time, with Orientation~3 tended to be slower and Orientation~4 faster relative to other orientations. In \textit{DuoZone}, no orientation pairs performed significantly different than others, consistent with a more uniform performance across orientations.

For Task 2, normality assumption was not violated for baseline ($p = 0.398$) and DuoZone($p = 0.939$) using Shapiro–Wilk test. A two-tailed paired-sample T-test shows that users in DZ-2 condition is significantly faster ($t(15) = 3.23, p = 0.006$) than that of the baseline, with DuoZone scores an average time of $168.3$ seconds ($SD = 67.5$) when user is done with the setup and the baseline scores an average of $218.1$ seconds ($SD = 73.1$). 

\begin{figure}[h]
    \centering
    \includegraphics[width=.9\linewidth]{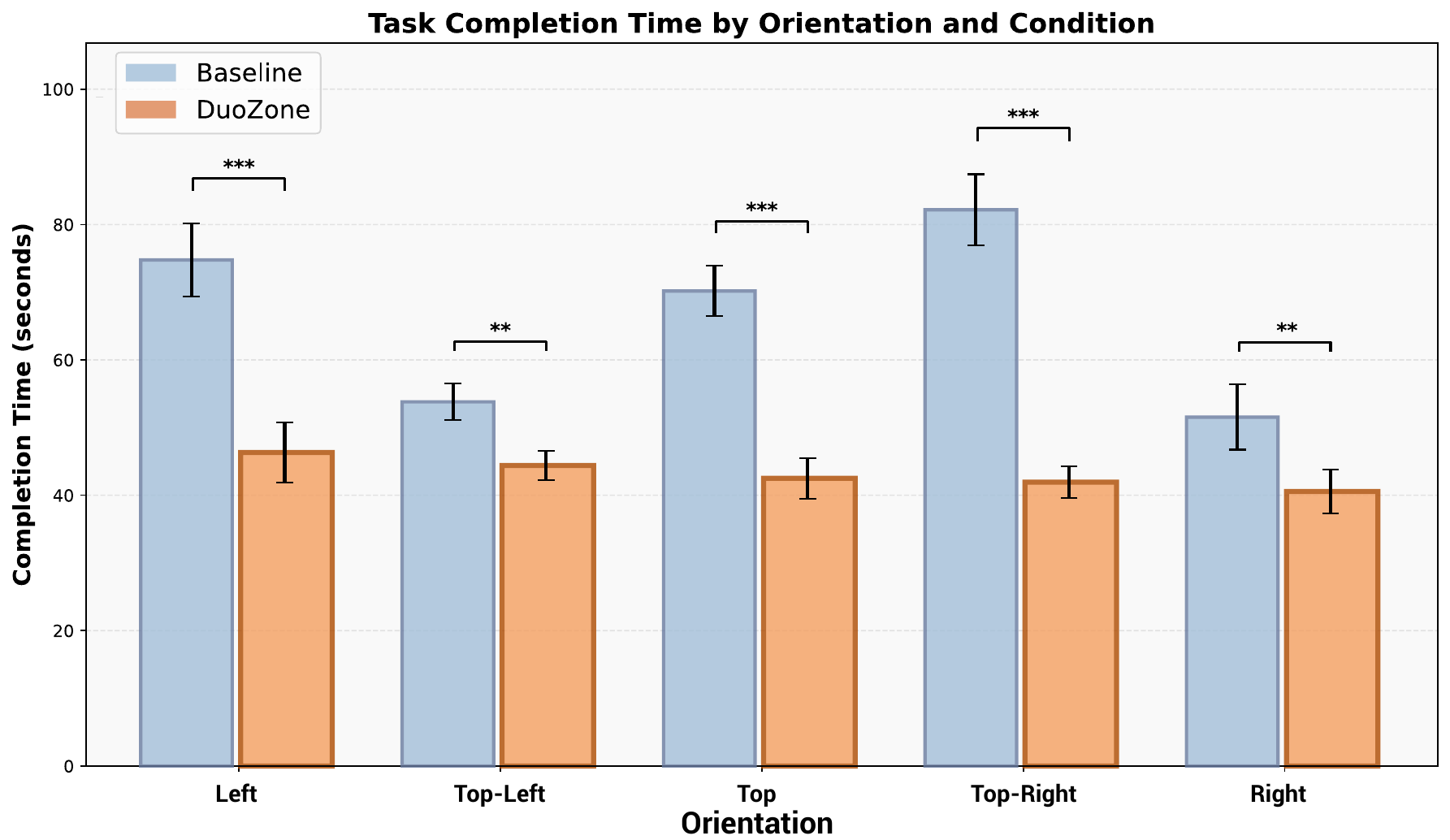}
    \caption{DuoZone significantly reduces completion time across orientations. Bars show mean task completion time (±SE) for Baseline vs. DuoZone at five screen orientations. In every orientation, DuoZone takes less time, and its times cluster tightly (~low-40s) while Baseline varies widely (~50–80s), indicating both speed gains and orientation-invariant performance. Brackets with asterisks mark significant between-condition differences for that orientation}
    \label{fig:task1_completion_time}
    \vspace{-8pt}
\end{figure}

\begin{figure}[h]
    \centering
    \includegraphics[width=0.9\linewidth]{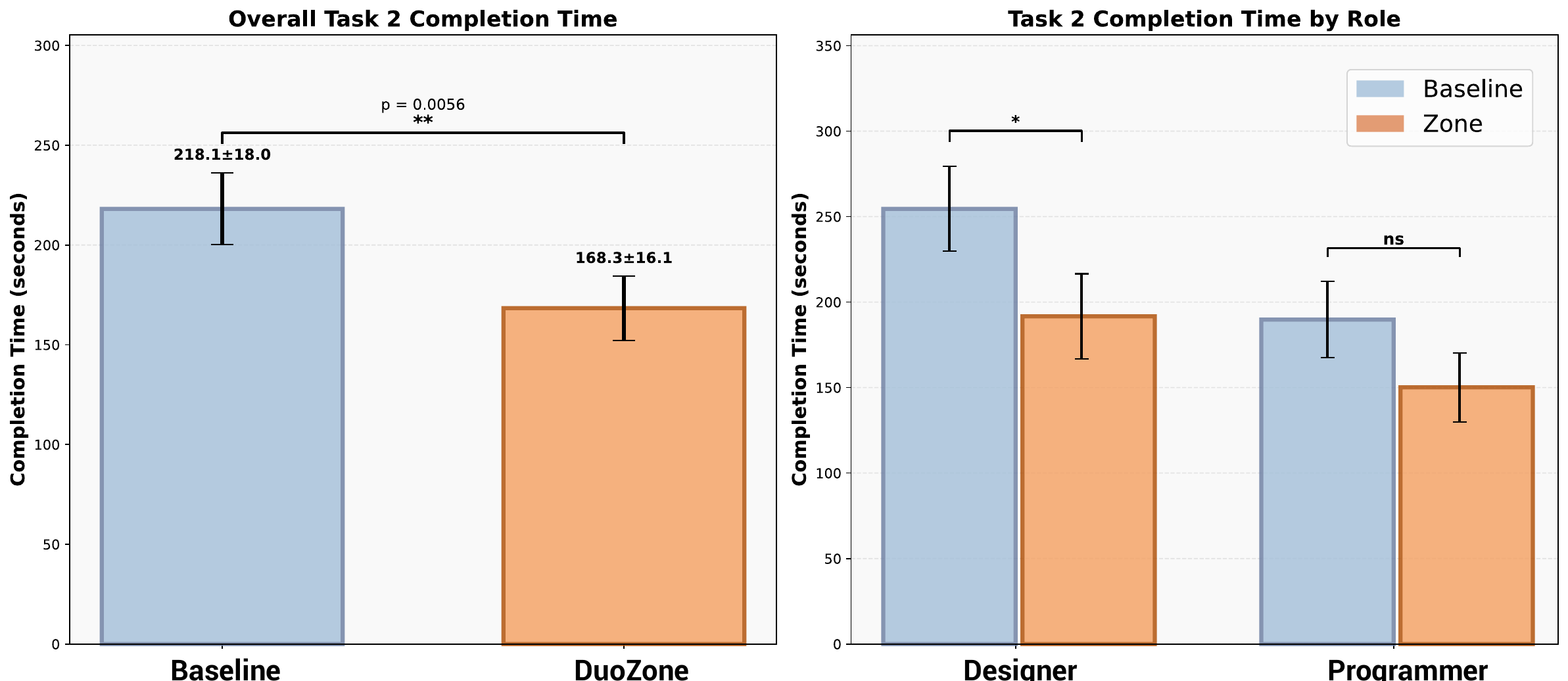}
    \caption{DuoZone shortens workspace construction time. The left charts show overall completion time for the workspace construction task. DuoZone (168.3 ± 16.1 s) is significantly faster than Baseline (218.1 ± 18.0 s, p = 0.0056). The right charts break results down by role: Designers show a significant speed improvement with DuoZone (p < 0.05), while Programmers also perform faster.
Together, the two plots indicate DuoZone consistently reduces workspace setup time across user types.}
    \label{fig:task2_completion_time}
        \vspace{-8pt}
\end{figure}

\subsection{Cognitive Load}
The TLX's subscale ratings did not meet the normality assumption using Shapiro--Wilk test~($p<0.05$); therefore nonparametric tests are 
used for this analysis. 

For Task 1, a Kruskal--Wallis test revealed a \textbf{significant difference in overall cognitive load}, $H(1)=7.57$, $p=.006$, with the \textit{DuoZone} condition ($M=33.30$) reporting lower scores than \textit{Baseline} ($M=47.71$).Post-hoc two-sided Mann--Whitney $U$ tests indicated that \textit{Effort} ($p=.002$, $r=.56$), \textit{Frustration} ($p=.010$, $r=.46$), and \textit{Physical Demand} ($p=.019$, $r=.42$) were all significantly reduced under \textit{DuoZone} relative to \textit{Baseline}; meanwhile, performance is significantly improved for DuoZone condition ($p=0.031, r=0.48$). No other subscales reached significance—\textit{Mental Demand} ($p=.250$), \textit{Temporal Demand} ($p=.186$), and \textit{Success} ($p=.135$)—though \textit{Success} tended to be higher for \textit{DuoZone}.

For Task 2, a Kruskal--Wallis test found a significant difference in the overall cognitive load ($H(1)=8.21$, $p=.004$), with \textit{DuoZone} condition($M=22.38$) less than \textit{Baseline} ($M=34.16$). A post-hoc pairwise comparison with two-sided Mann--Whiteney U test further found that \textit{Mental Demand}($p=0.017$) and Effort ($p=0.029$) showed significantly lower rating in \textit{DuoZone} conditions than that of the \textit{Baseline} with medium effects ($r = 39$ and $.42$). No significant differences were identified from the rest of the subscales. 

\begin{figure}[h]
    \centering
    \includegraphics[width=0.9\linewidth]{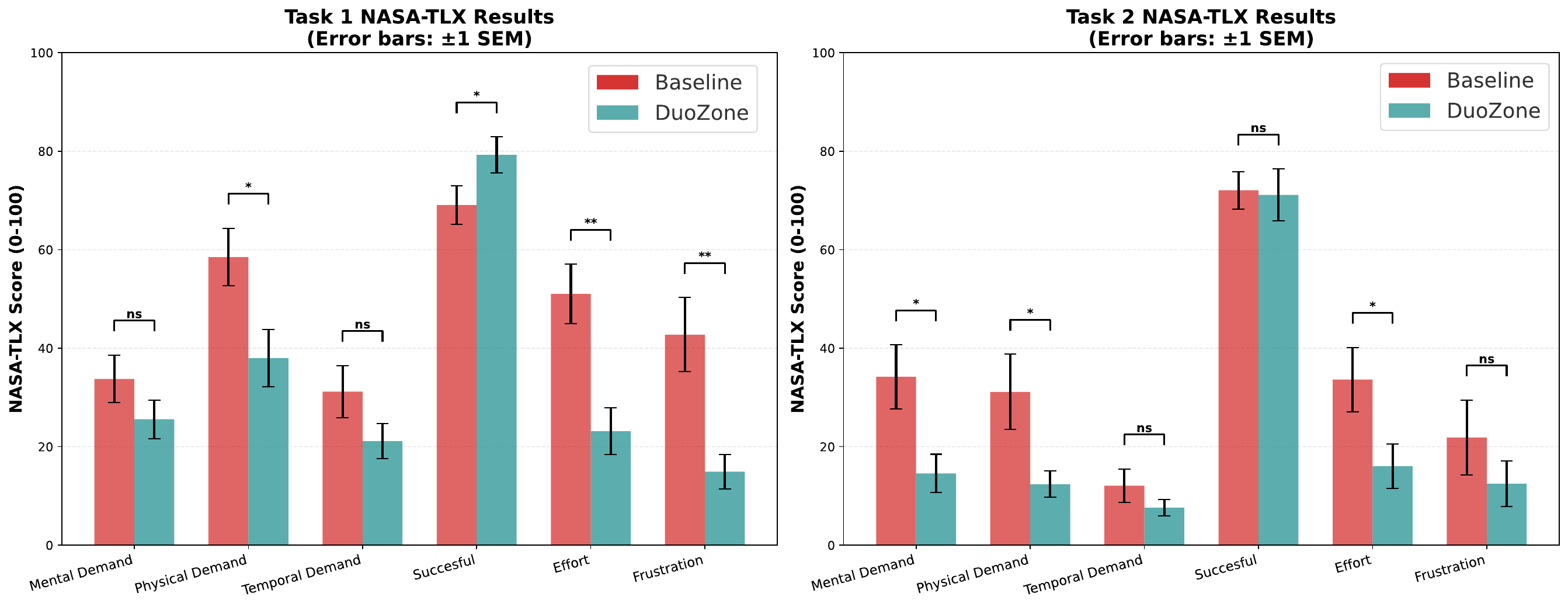}
    \caption{Comparison of NASA-TLX subscale scores between Baseline and DuoZone conditions for both Task 1 and Task 2. DuoZone consistently shows lower ratings in physical, mental, and effort-related demands, indicating reduced cognitive load and frustration, while maintaining or improving perceived performance compared to the manual Baseline.}
    \label{fig:tlx_comparison}
        \vspace{-8pt}
\end{figure}

\subsection{Acceptance of AI recommendations}
\textbf{High-level Goal related application recommendation: } On average, each user created about 4 zones, with AI recommended 8 virtual apps per each workspace setup. Within these recommendations, users rejected on average of 0.68 apps ($\sigma = 0.87$), resulting in an average of \textit{90.3$\%$} acceptance rate ($\sigma=13.7\%$). 

\textbf{Cost-model driven layout sizing and ordering} After the cost-model adjusts the layout of the zones, the average user further adjusted 0.68 out of 4 zones' layout, or accepting $82.8\%$ of AI layout suggestion. In terms of ordering and adjacency, the average user accepted $76.5\%$ of automatic ordering. Meanwhile, manual layout scaling accounts for $39.6\%$ or ($60.4\%$ acceptance rate).

\begin{figure}[h]
    \centering
    \includegraphics[width=1.0\linewidth]{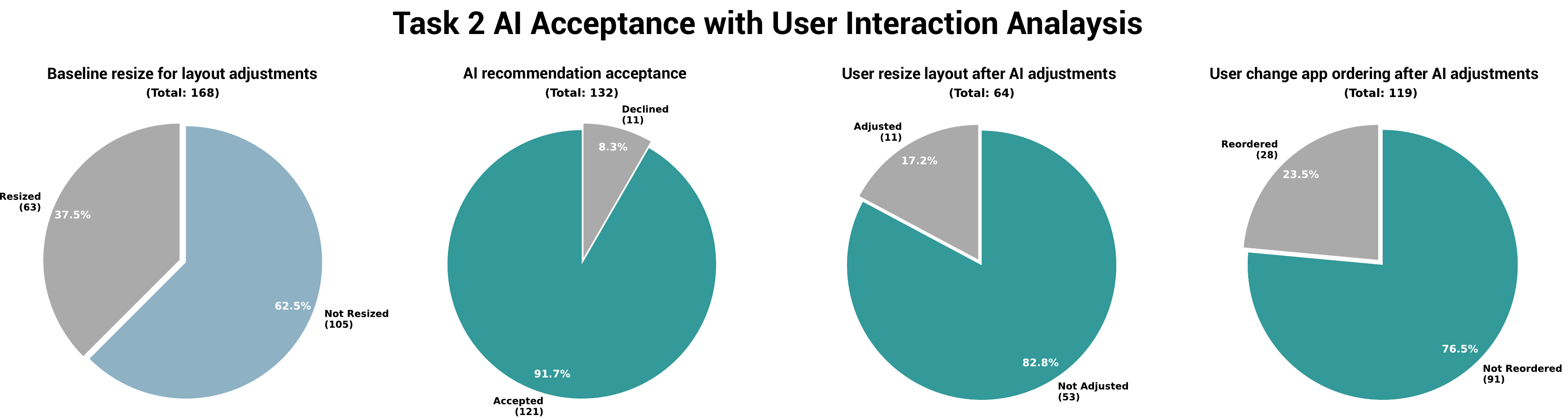}
    \caption{Visualization of user interaction and AI recommendation acceptance in Task 2. The charts illustrate how users accepted or adjusted AI-generated workspace layouts, showing high acceptance of recommended applications and layouts, with relatively few post-adjustments in resizing or reordering compared to the manual baseline.}
    \label{fig:AI_analysis}
        \vspace{-8pt}
\end{figure}

\subsection{Qualitative Analysis}
We coded the survey and interview data using both open and axial coding. Two coders separately coded the data followed by a cross-check to enhance reliability. 

\subsubsection{Preferences are dependent on users' need for precision or stability.}
Preferences for the baseline or arrangement zone depends on the level of controls and how ``spontaneous'' participants want for the task. Baseline was preferred when precision is required or when on-the-fly and quick adjustments are needed (P9, P12). It is also easy to quickly zoom in and out on a single virtual window for design situations (P13). Moreover, participants appreciated independent control over each virtual window as the adjustment in baseline condition does not affect others (P6, P15, P16), or for highly customized layouts (P8, P16).

Preference for arrangement zones are more towards stability, organization, and consistency (P1, P4, P6, P8, P10, P11, P12), as many participants brought up that the zones are suitable for daily multitasking and for specific tasks with consistent layouts (P3, P5, P9, P14, P15). In addition, several participants reported a preferred two-stage strategy: completing initial setup using DuoZone's arrangement zones, drop in virtual windows and manually fine-tune the zone's layout (P7, P9, P12, P10, P11). Overall, the structured DuoZone reduces setup costs while the Baseline maintains advantages in precision and control.

\subsubsection{Participants feedback on AI acceptance}
Overall, participants reported similar satisfaction ($W=73.50$, $p=.927$) for Manual ($M=3.81$, $SD=0.88$) and DuoZone ($M=3.69$, $SD=1.06$) conditions.

\textbf{Recommendations create usable, intention-aligned layouts.} Overall, participants are positive on the virtual applications recommended to them due to both high relevance (P9, P11, P13, P15, P16) and reasonable workspace layout and sizing. P11 found that the recommended apps ``perfectly align with my intention...exactly what I want'', and P3 and P15 affirmed that AI's coarse layout ``matches my thinking'' and is great for quickly getting a usable scaffold when they ``have no idea''. Even when unnecessary apps were included, participants felt the necessary components were present (P13, P15, P16). 

\textbf{Primary apps are well-centered and other clustered apps support streamlining workflow. } Users highly valued that AI place the right app in the center of their vision (P7, P8, P16), for example IDE for coding or a design tool. Participants also noted that the AI placed related supporting applications near each other, which facilitates the workflow and reduces searching effort. P7 said that ``...placing File and Note together makes it easier to access and view''; ``Gmail next to a chat is useful because both information sources required for decision-making are visible simultaneously.'' (P8).

\textbf{AI layouts are well-received but could benefit from personalized fine-tuning} Participant (P1) also noted that ``AI doesn't know my personal work habits...working is very personalized'' and P7 said that ``without my prior experience as input, it's hard to infer...''. P16 added that ``AI gives the most of size relevant, but I need to adjust the details''. These findings suggested that participants accept the general structure but still need to finetune the app's size to match their habits. In addition, participants tend to adjust zones wider horizontally and flatter vertically. As P10 brought up ``Photoshop is too narrow", P15 brought up ``A PDF reader can't be that narrow'',and P6 said that ``A squashed browser is hard to use...''. These indicates that sizing specific sizing requirement maybe different for each application and its use. The overall text readability was good, as P13 noted that ``I can see all these words clearly''.

\subsubsection{User-created affordance: AI scaffolds, user perfects}
Despite imperfect sizing, ordering and app placement, participants are generally positive towards a collaborative approach. P3 said that ``AI's layout matches my thinking, but some sizes aren't perfect;I can fine-tune on that framework'', while P12 mentioned that AI can ``just give me a framework, then i refine what i truly need''. P15 further brought up that `` AI does the layout fast; I'll modify afterward. This is convenient.''. Overall, acceptance is high on the framework-plus-selective-adjustments approach to retain control over critical details. 

\subsubsection{Manual adjustment enables precise, flexible layouts but is time-consuming and scales poorly, causing crowding and self-occlusion.}
Participants praised the the high-level of customization and flexibility in manually adjusting the virtual windows, with the ability to adjust individual apps freely(P5, P8, P16). Like DuoZone, participants reported that manual placement allow them to precisely order applications with relevance and use frequency (P2, P8), and considered this a suitable to design or program. Manual adjustment also allow them to precisely place supplemental materials such as documents, image assets library, or notes to the side for easy access (P1, P7, P15).

The most downside to the manual adjustment is its operational complexity and lack of efficiency. Participants mentioned that they had to ``manually drag and adjust every application'' and feels in a ``time consuming process''(P5).  Moreover, as more applications pile into the XR space, participants reported space constraints and crowding despite XR's virtually infinite interaction space (P15). Several participants also reported overlapping issues when trying to form layouts with manual adjustment, which further causes selection and visual occlusion issues (P15, P15).

\subsubsection{Occlusion is vital for awareness, reduced distraction, multitasking, and agency}
Participants described the occlusion as an essential element of effective and trustworthy XR interaction (P13, P14, P15, and P16). Participants emphasized occlusion Zone's partial transparency sustains peripheral awareness without disrupting focus (P15). For example, they would “place it on the desk view to see my hands and computer” and “make sure I can see if a colleague comes over”. P16 extended this logic, explaining that the feature should “activate when someone approaches or when I don’t want to be disturbed,” and proposed lightweight cues such as “a flashing screen edge or glowing text when someone is near.” P15 framed occlusion as critical for both social and practical reasons: “If my boss walks in while I’m slacking, I need it near the door,” while also wanting to “see my paper, pen, and phone notifications” within their workspace. P16 highlighted the experiential value of maintaining environmental connection, stating that occlusion “lets me know when someone greets me” and “allows me to enjoy the view while still working.” When considering varied contexts, participants described the usefulness of occlusion zones for context-sensitive adaptation: in cafés, P15 would adapt “a clear center area to see the cup and desk,” whereas P16 emphasized that in kitchens “the center should remain transparent to identify obstacles,” suggesting transparency “briefly triggered by sound or movement.” P16 valued the ability to “see books or drawers when reading or designing,” and P16 envisioned using occlusion “to track cooking timers while reading at the dining table.” Collectively, participants articulated a coherent design vision in which dynamic, gaze- or sound-triggered occlusion operates as a core mechanism for maintaining situational awareness, supporting multitasking, and reinforcing user agency in extended reality workspaces.

%% file: sections/05_discussion.tex
\section{Discussion}


\subsection{DuoZone Improve Interaction Speed and Reduce Cognitive Load}
We found that DuoZone significantly improved temporal efficiency and reduced cognitive workload (RQ1). In Task 1, completion time decreased by 34.5\% compared with the manual baseline (p < .001), indicating that structured Arrangement Zones accelerate micro-operations such as dragging, snapping, and resizing. Task 2 (workspace construction) also showed shorter setup times (p = .006), revealing that the mixed-initiative condition benefits both fine-grained and holistic layout work. Cognitive load (NASA-TLX) dropped across physical, effort, and frustration subscales, with DuoZone reducing median workload scores by roughly 1/3.

Participants’ accounts reinforce these findings. They described DuoZone as promoting ``structure,'' ``stability,'' and ``consistency,'' which reduced the need to constantly monitor or adjust spatial relations. Rather than fragmenting attention between spatial arrangement and task content, users could focus on their primary goals. The structured layout mitigated physical fatigue from frequent controller gestures, while the AI-assisted recommendations reduced the sense of cognitive clutter. In this sense, DuoZone acts as both a cognitive offloading mechanism and a spatial anchor, helping users maintain mental orientation during complex multitasking.

\subsection{DuoZone Reduce Manual Adjustments to Virtual Windows}
DuoZone demonstrably minimized the need for extensive manual repositioning (RQ2). High acceptance rates of 82.8\% for layout structure and 76.5\% for ordering show that most AI-generated configurations were perceived as immediately usable. The remaining manual resizing (less than 40\%) indicates that while automation is highly effective at producing viable initial configurations, personal nuances in workspace aesthetics or application preferences still prompt minor refinements. 

The qualitative data reveals that DuoZone reduces the number of initial, complex manual adjustments by providing an accepted framework. Participants noted that manual adjustment is operationally complex and requires users to ``manually drag and adjust every application''. Interviews suggest that users conceptualize DuoZone’s AI output as a “scaffold rather than a final product.” They trust the AI to handle the “heavy lifting” of structural organization, freeing them to make expressive adjustments.  They often employ a ``two-stage mental model'' where ``AI scaffolds, user perfects''. Users accepted the general structure to personally match personal work habits. This indicates that our system reduces the initial interaction cost of arrangement but preserves manual control for granular details
\subsection{DuoZone Provide Useful Recommendations}
DuoZone’s AI recommendations were not only accepted at high rates but also meaningfully aligned with user intentions (RQ3). The 90.3\% application relevance score and 82.8\% layout acceptance rate reveal a robust ability to infer contextual needs. The consistency in adjustment direction further suggests that users respond predictably to structured but flexible recommendations. This tailors them to specific application affordances rather than rejecting them wholesale.

Participants praised the AI for its ability to align perfectly with user intent and facilitate workflow. Users often found the recommended applications highly relevant. Even when unexpected apps were recommended, participants felt the necessary components were present. They commonly admitted DuoZone provided a great way to ``quickly getting a usable scaffold'' when they ``have no idea.'' Meanwhile, participants highly valued the strategic placement of windows. The AI placed the main work application (e.g., IDE for programming and Canva for designing) in the center of their vision. Furthermore, related supporting applications were constantly placed near each other to facilitate workflow and reducing searching effort. 

\subsection{Comparison Between DuoZone and Manual Setup}
DuoZone achieved significantly higher efficiency with equivalent user satisfaction (RQ4). The system cut setup time by about 50 seconds and reduced both Mental Demand and Effort, yet users rated satisfaction on par with manual methods. This parity implies that users did not perceive automation as compromising quality or expressiveness. Interestingly, the similar rates of fine-tuning across conditions (39.6\% vs. 37.3\%) suggest that DuoZone preserves natural user behavior while expediting the initial layout phase.

This balance reflects a deeper trade-off between flexibility and structure. Manual setup affords maximum freedom but imposes heavy operational and cognitive costs. Users described it as “time-consuming” and prone to visual clutter and occlusion. DuoZone, in contrast, imposes light structural constraints that users found stabilizing rather than restrictive. The AI’s involvement does not replace manual creativity but reframes it from construction to refinement. As a result, DuoZone bridges the gap between the precision of manual arrangement and the convenience of automation. We have proven our system embodies a mixed-initiative collaboration model that enhances both usability and user control.

\subsection{Scalability: General Scenes vs. Task-Specific Workspaces}

LLM-driven Recommendation Zone extend scalability through context-aware reasoning, translating semantic intent (e.g., ``coding a web game'') into actionable layouts that pre-structure the workspace. Participants valued this as goal-aware scaffolding, describing the system from ``usable starting point'' to ``perfectly aligns with my intention.'' The automation thus reduces initial friction while preserving control. Also, users (P11,15) find it useful in one-off, low-frequency activities where they lack practiced routines (e.g., ``cook rice''), so DuoZone can quickly stage all steps and resources. Participants (P12, 16) also emphasized the importance in organizational settings where workers may lack personal machines and tool chains differ across companies. For example, some company may prefer ``Gmail over Outlook'' or  ``Slack over Discord.'' Especially during early on-boarding to a new company, AI-orchestrated workspace setup helps users rapidly internalize local application norms and instantiate enterprise-preferred configurations, delivering quick hands-on familiarity and lowering cross-tool friction. When a stable ``home base'' is unavailable, this AI mediation becomes the most convenient, and often the only practical means to get productive fast.

\subsection{Agency and Trust in Mixed-Initiative Systems}

User reflections reveal that DuoZone’s architecture successfully cultivates both agency and trust by positioning automation as a collaborative partner rather than a replacement for human control. Participants consistently adopted a human-AI collaborative framework where the system generates a usable structural foundation that users refine to their personal preferences. The Arrangement Zone amplifies this idea by allowing users to deliberately plan spatial layouts and maintain a sense of authorship over their XR environments. Trust is strengthened through goal alignment and strategic spatial reasoning. The Recommendation Zone was repeatedly described as ``perfectly aligned with my intention,'' with users appreciating how the AI placed primary tools centrally and grouped related applications proximally (e.g., ``File and Note together''). These intelligent placements reduced visual search and improved workflow coherence, leading users to describe the AI as ``reliable'' and ``thoughtful.'' 


%% file: sections/06_limitation_n_future.tex
\section{Limitation}
Despite the benefit from DuoZone, the work is limited in the following ways. First, its reliance on LLM makes it difficult for offline use, and the performance can be affected by the internet communication and the stability of a remote LLM. Second, AI recommendation lacks a deeper understanding of users' work habits. Future work can explore a human-in-the-loop update cycle for the AI system to keep track on users' physical status, salient signals, and interactions to offer more in-depth personalization. Third, occlusion-free areas are evaluated qualitatively as it is not the main focus of this work, but used as a compatible extension of the Arrangement Zone.Future work could test long-term interaction using fully functional XR apps and dive into their effects in real-life scenarios.. 

%% file: sections/07_conclusion.tex
\section{Conclusion and Limitation}
We present DuoZone, a human-AI collaborative XR window management to improve the interaction performance and lowering cognitive load over current window management practice. The system established a mixed-initiative interaction scheme via two types of spatial configurations (zones) with six templates that allows users to gain efficiency while retaining full control. The \textbf{Arrangement Zone} enables spatial anchors in the XR space for swift snapping, resizing, and group positioning; and \textbf{Recommendation Zone} uses users' high-level goals and an interaction cost model to automatically recommend apps needed and to adjust the layout and sizing for better experience and less effort. A sixteen user empirical study found that DuoZone significantly reduces users' cognitive load and completion time, while AI-recommended apps have over 90\% of acceptance rate. The work demonstrated a balanced window management with automation and agency and provide an easy to use and less effort way for XR workspace productivity.

%% file: sections/08_acknowledgement.tex
\section{Acknowledgement}
We thank Shirley Hu for helping and ideating the teaser figure of the work. Generative AI (ChatGPT 5 and Gemini) is used improve writing and grammar corrections. 